\documentclass[showpacs,10pt,twocolumn,prb]{revtex4-1}

\usepackage{amsmath}
\usepackage{amssymb}
\usepackage{graphicx}
\usepackage{amssymb}
\usepackage{graphics}
\usepackage{epsfig}
\usepackage{CJK}
\usepackage{color}

\begin{document}

\title{Electron-hole asymmetry, Dirac fermions and quantum magnetoresistance in BaMnBi$_{2}$}
\author{Lijun Li,$^{1\dag}$ Kefeng Wang,$^{1,\ast}$, D. Graf,$^{2}$ Limin Wang,$^{1,\ast}$ Aifeng Wang$^{1}$ and C. Petrovic$^{1}$}
\affiliation{$^{1}$Condensed Matter Physics and Materials Science Department, Brookhaven National Laboratory, Upton, New York 11973, USA\\
$^{2}$National High Magnetic Field Laboratory, Florida State University, Tallahassee, Florida 32306-4005, USA}

\date{\today}

\begin{abstract}
We report two-dimensional quantum transport and Dirac fermions in BaMnBi$_{2}$ single crystals. BaMnBi$_{2}$ is a layered bad metal with highly anisotropic conductivity and magnetic order below 290 K. Magnetotransport properties, nonzero Berry phase, small cyclotron mass and the first-principles band structure calculations indicate the presence of Dirac fermions in Bi square nets. Quantum oscillations in the Hall channel suggest the presence of both electron and hole pockets whereas Dirac and parabolic states coexist at the Fermi level.
\end{abstract}
\maketitle

\section{Introduction}

In Dirac materials such as graphene,topological insulators Bi$_{2}$Se$_{3}$ and Bi$_{2}$Te$_{3}$, copper oxides and BaFe$_{2}$As$_{2}$ iron pnictide the energy spectrum of low-energy electrons can be approximated by the relativistic Dirac equation.\cite{Dirac,NovoselovK,RMP 82 3045, RMP 83 1057,Science 288 468,PRL 104 137001,NM 6 183, PRL 53 2449} The effective Hamiltonian is characterized by Pauli matrices and even for nonzero effective mass the eigenstates of the Dirac hamiltonian are spinor wavefunctions. One of the most interesting aspects of Dirac fermions is quantum transport phenomena.\cite{RMP 81 109,NC 1 47, PRB 58 2788} In contrast to the conventional electron gas with parabolic energy dispersion, the distance between the lowest and first LLs of Dirac fermions in a magnetic field is large and the quantum limit where all carriers are condensed to the lowest LL is easily realized in moderate magnetic fields.\cite{Fundamentals,Nature 438 201, Science 324 924,PRL 106 217004} Consequently, quantum Hall effect and large linear magnetoresistance (MR) could be observed.

\begin{figure}
\centerline{\includegraphics[scale=0.37]{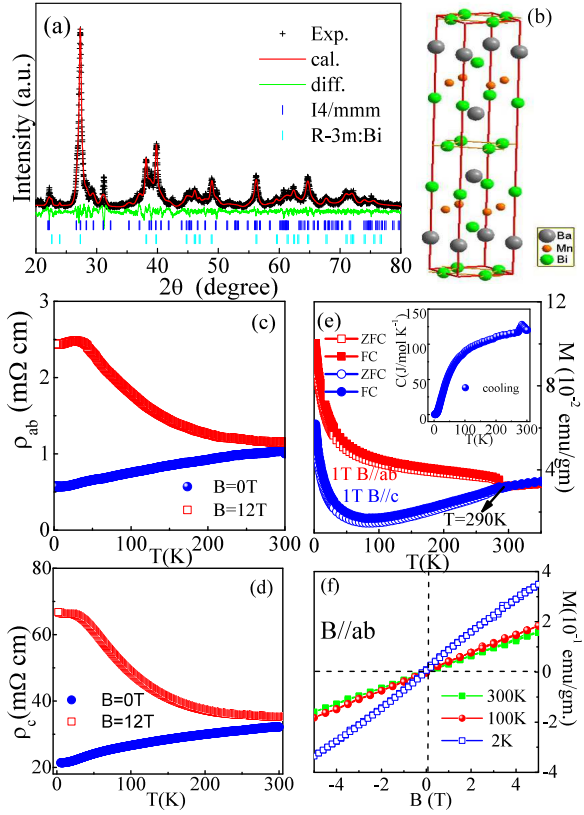}} \vspace*{-0.3cm}
\caption{(Color online) (a) Powder XRD patterns and refinement results. The data were shown by (+), fitting and difference curves are given by the red and green solid line respectively. (b) Crystal structure of BaMnBi$_{2}$. (c) and (d) Temperature dependence of the in-plane resistivity $\rho_{ab}(T)$ and $c$-axis resistivity $\rho_{c}(T)$ of the BaMnBi$_{2}$ single crystal in the $B$ = 0 T (solid symbols) and $B$ = 12 T (open symbols) magnetic fields, respectively. (e) Temperature dependence of magnetization $M (T)$ for applied magnetic field parallel and perpendicular to the $ab$ plane in both zero-field-cooling (ZFC, solid symbols) and field-cooling (FC, open symbols). The inset in figure 1(e) shows temperature dependence of heat capacity. (f) Magnetization hysteresis loops M(H) of BaMnBi$_{2}$, at 300 K, 100 K, and 2 K.}
\label{XRD}
\end{figure}

Besides two-dimensional (2D) electronic states in materials such as graphene and topological insulators, bulk AMnBi$_{2}$ crystals (A = alkaline earth metals such as Sr or Ca) show quasi-two-dimensional (quasi-2D) quantum transport of Dirac fermions.\cite{PRB 84 220401,WangKF2,PRL 107 126402,JiaLL} Their crystal structure features alternating layers of Mn-Bi edge-sharing tetrahedra and a Bi square net, separated by a layer of A metals. Square bismuth nets host two dimensional Dirac dispersion in the bulk band structure.\cite{PRB 84 220401,WangKF2,PRL 107 126402,JiaLL,PRB 87 245104,PRB 84 064428} The characteristics of the Dirac cone is governed by spin-orbit coupling and local arrangement of A atoms that surround the Bi square net. For A=Sr the degeneracy along the band crossing line is lifted except at the place of anisotropic Dirac cone. This is in contrast to A=Ca where the energy eigenvalue difference due to perturbation potential created by staggered Ca atomic layers cancels out, resulting in a zero-energy line in momentum space.\cite{PRB 87 245104,sr 4 5385} Therefore, the selection of A atoms in AMnBi$_{2}$ presents an opportunity to tune the contribution of Dirac states at the Fermi level, Dirac cone anisotropy and structure. Here we report quasi-2D quantum transport and Dirac fermions in BaMnBi$_{2}$. We show compensated nature of electronic transport and the presence of both Dirac-like and trivial carriers at the Fermi level.

\begin{figure}
\centerline{\includegraphics[scale=0.30]{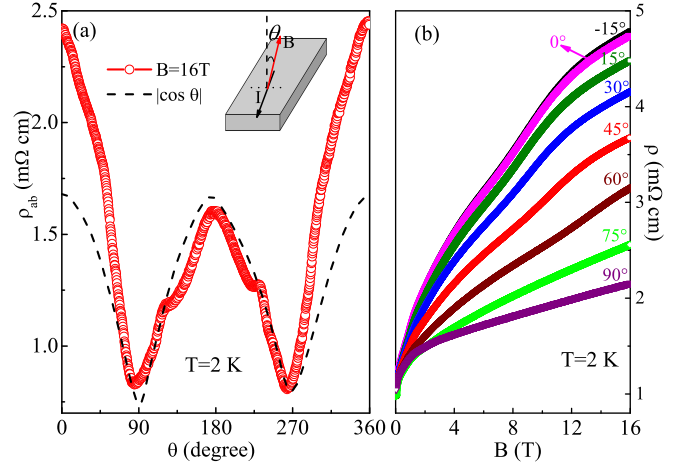}} \vspace*{-0.3cm}
\caption{Color online) The crystal was fixed on a rotating stage so that the tilt angle $\theta$ between the crystal surface ($ab$ plane) and the magnetic field can be continuously changed and the current is flowing in the $ab$ plane perpendicular to magnetic field. (a) Tilt angle $\theta$ (from 0$^{\circ}$ to 360$^{\circ}$) dependence of in-plane resistivity ($\rho_{ab}$) at $B$ = 16 T as $T$ = 2 K. The black dash line is the $|$$\cos(\theta)$$|$ curve. The inset shows the configuration of the measurement. (b) In-plane resistivity $\rho_{ab}$ vs magnetic field B at different tilt angles $\theta$ at 2 K (-15$^{\circ}$ to 90$^{\circ}$).}
\label{XRD}
\end{figure}

\section{Experiment}

Single crystals of BaMnBi$_{2}$ were grown from molten metallic fluxes.\cite{Fisk} X-ray diffraction (XRD) data were obtained by using Cu K$_{\alpha}$ ($\lambda = 0.15418 nm$) radiation of a Rigaku Miniflex powder diffractometer on crushed crystals. Electrical transport, magnetization and heat capacity measurements were performed in a Quantum Design PPMS and MPMS on cleaved and polished single crystals. Polishing is necessary in order to remove residual bismuth droplets from the surface of as-grown single crystals [Fig. 1(a)]. High-field magnetoresistance (MR) was performed at the National High Magnetic Field Laboratory in the same configuration as the in-plane MR. First principle electronic structure calculations were performed using experimental lattice parameters within the full-potential linearized augmented plane wave (LAPW) method implemented in WIEN2k package.\cite{Weinert,Blaha} The general gradient approximation (GGA) was used for exchange-correlation potential.\cite{Perdew} In the calculation, we adopted the antiferromagnetically stacked magnetic structure of SrMnBi$_{2}$.\cite{GuoYF}

\section{Results and Discussion}

The unit cell of BaMnBi$_{2}$ crystals can be indexed in the I4$/$mmm space group by RIETICA software [Fig. 1(a)].\cite{R 1998} The lattice parameters $a = b = 0.4627(8) nm$ and $c = 2.4315(7) nm$ are in agreement with the previously reported values.\cite{ZFN B 32 383} Hence, BaMnBi$_{2}$ crystal structure [Fig. 1(b)] features identical space group but somewhat elongated $c$-axis when compared to SrMnBi$_{2}$. The temperature dependence of in-plane resistivity $\rho_{ab}(T)$ and out of plane resistivity $\rho_{c}(T)$ [Fig. 1(c) and Fig. 1(d), respectively] suggest considerable transport anisotropy with $\rho_{c}(T)$/$\rho_{ab}(T)$ $\approx$ 20. Electrical resistivity implies a bad metal with residual resistivities $\rho_{ab0}$ $\approx$ 0.57 m$\Omega$cm and $\rho_{c0}$ $\approx$ 19.36 m$\Omega$cm.  In both CaMnBi$_{2}$ and SrMnBi$_{2}$ $\rho_{c}$(T) features a broad maximum below 300 K, usually interpreted as a high-T incoherent to low-T coherent conduction in quasi-2D electronic systems.\cite{Gutman1,Gutman2} The temperature dependence of c-axis resistivity [Fig. 1(d)] suggests that BaMnBi$_{2}$ E$_{F}$$\tau$ values (where E$_{F}$ is Fermi energy and $\tau$ is the scattering time) are different from CaMnBi$_{2}$ and SrMnBi$_{2}$ so that the coherent-incoherent crossover is above 300 K. An external magnetic field $B$ = 12 T strongly enhances the resistivity. The MR ratio MR = [$\rho(B)-\rho(0)]/\rho$(0) exceeds 300\% for the in plane MR (MR$_{ab}$) and 200\% for the $c$-direction MR (MR$_{c}$) at 2 K and 12 T. Temperature dependence of the magnetic susceptibility $M(T)$ measured in 1 T is shown in Figure 1 (e). The $M-T$ curves are similar to SrMnBi$_{2}$ and CaMnBi$_{2}$ with a magnetic anomaly around 290 K.\cite{WangKF2,PRL 107 126402,APL 100 112405,PRB 84 064428} A clear $\lambda$-type peak appears in heat capacity at around $T$ = 290 K whereas the linear specific heat coefficient $C$$\sim$$\gamma$$T$ is obtained to be 34.7(1) mJ mol$^{-1}$ K$^{-2}$ from the specific heat data [Fig. 1(e) inset]. Magnetization hysteresis loops [Fig. 1(f)] also support the antiferromagnetic nature of $T$ = 290 K transition.

Figure 2(a,b) shows field-angular-dependence of the in plane resistivity $\rho_{ab}$($B$, $\theta$) at $T$ = 2 K.  When $B$ is parallel to the $c$ axis ($\theta$ = 0$^{\circ}$, 180$^{\circ}$, 360$^{\circ}$), the MR is maximized and is linear in field for high fields with the presence of angular - dependent quantum oscillations. With increase in the tilt angle $\theta$, the MR gradually decreases to the minimal value when $B$ is in the $ab$ plane ($\theta$ = 90$^{\circ}$, 270$^{\circ}$). The MR bears close resemblance to the $|$$\cos(\theta)$$|$ angular dependence [dashed black line in Fig. 2(a)] observed in SrMnBi$_{2}$ and CaMnBi$_{2}$,\cite{PRB 84 220401,WangKF2} however with some deviations. This could suggest the presence of 3D, in addition to quasi-2D electronic transport in BaMnBi$_{2}$.

\begin{figure}
\centerline{\includegraphics[scale=0.2]{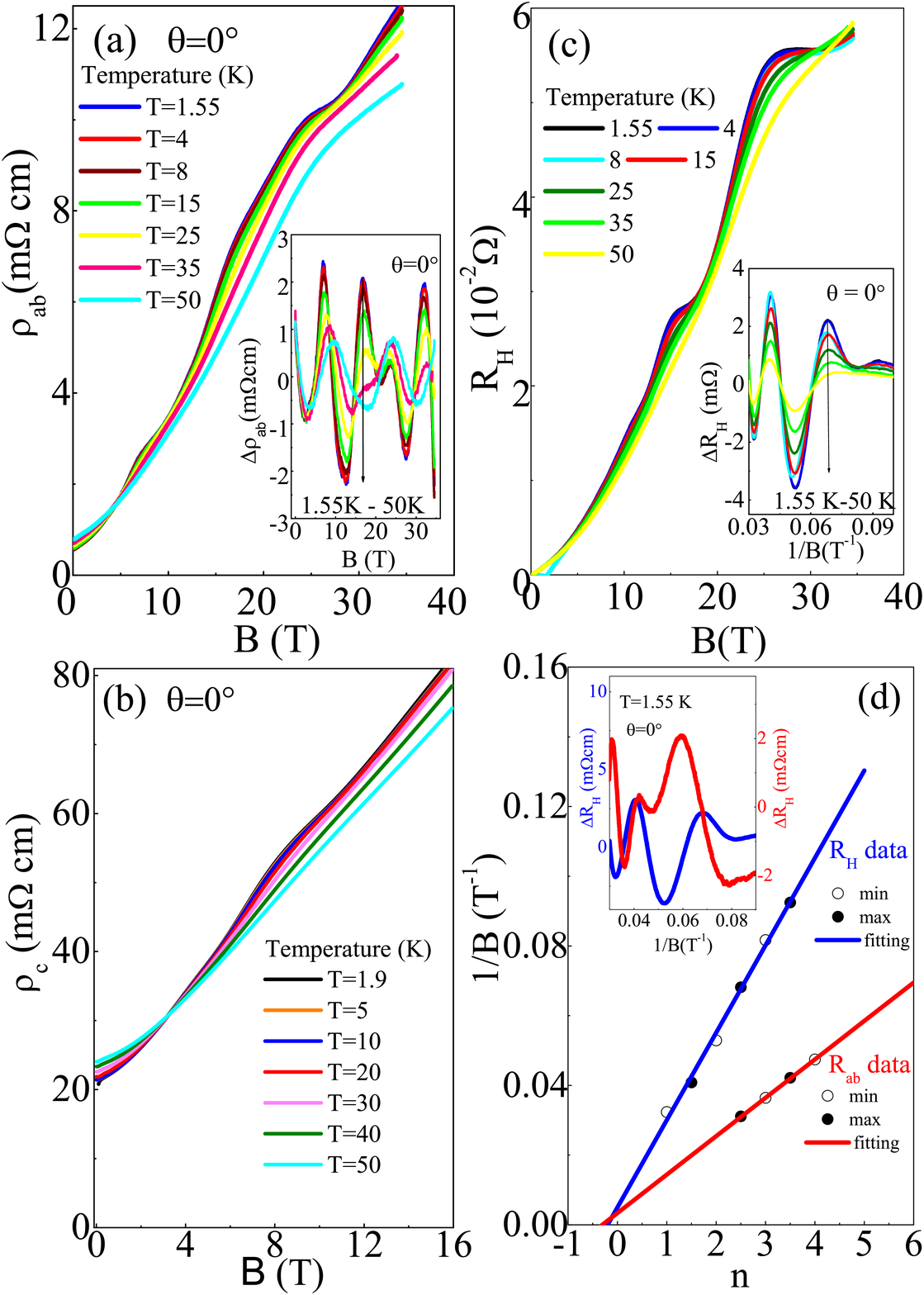}} \vspace*{-0.3cm}
\caption{(Color online) (a) In plane $\rho_{ab}-B$ and (b) $c$-axis resistivity $\rho_{c}-B$ of BaMnBi$_{2}$ crystal measured at different temperatures. (c) Magnetic field dependence of Hall resistance. (d) The integer Landau levels as a function of inverse field; the arrows indicate the positions of the estimated LL index $n$ labeled by the numbers. Insets in (a) and (c) present magnetoresistant SdH oscillations of $\Delta$$R_{i}$ = $R_{i}$ - $\langle$$R_{i}$$\rangle$ as a function of B where i=ab,c.}
\label{XRD}
\end{figure}

\begin{figure}
\centerline{\includegraphics[scale=0.4]{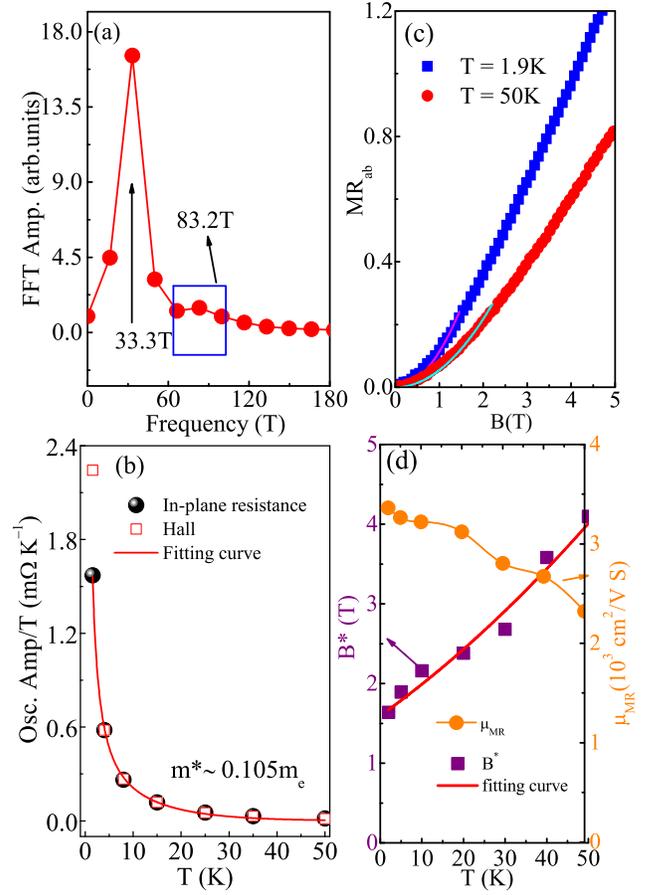}} \vspace*{-0.3cm}
\caption{Color online) (a) The Fourier transform spectrum of the SdH oscillation at $T$ = 1.55 K. (b) Temperature dependence of the temperature divided oscillation amplitude (Osc.Amp./$T$-$T$) in magnetoresistant SdH oscillations, the black balls are obtained from $\Delta$$R_{ab}$ data, and the red squares are obtained from $\Delta$$R_{H}$ data. The red line is the fitting results giving cyclotron mass $m^{\ast} \sim 0.105m_{e}$. (c) In-plane low-field magnetoresistance ($MR_{ab}$) for $T$ = 1.9 K (blue solid squares) and $T$ = 50 K (red solid circles); the lines are fitting results of $MR$ = $A_{2}$$\times$$B^{2}$.(d) Temperature dependence of the critical field $B^{\ast}$ (purple solid squares) and the effective $MR$ mobility (orange solid circles) $\mu_{MR}$ obtained by fitting the low-field $MR_{ab}$. The red solid line is the fitting results of $B^{\ast}$ using: $B^{\ast}=\frac{1}{2e\hbar\nu_{F}^{2}}(E_{F}+\kappa_{B}T)^{2}$.}
\label{XRD}
\end{figure}

Large anisotropy can be observed in BaMnBi$_{2}$ resistivity in magnetic field [Fig. 3(a,b)] when the current is applied along a and c axes. Inset in Fig. 3(a) shows magnetic field dependence of the $\Delta$$R$$_{ab}$ = $R_{ab}$-$\langle$$R _{ab}$$\rangle$ curves, where $\langle$$R_{ab}$$\rangle$ is a smooth background. With the increase in magnetic field, oscillations appear at fairly low magnetic fields at low temperatures. Magnetic field dependence of Hall magnetoresistance $R_{H}(B)$ also shows quantum oscillations below 50 K [Fig. 3(c)]. This is clear from the inverse field dependence of $\Delta$$R_{H}$ = $R_{H}$-$\langle$$R_{H}$$\rangle$ [Fig. 3(c) inset]. The oscillating component shows a periodic behavior in $1/B$. Figure 3(d) presents Landau levels (LL) index versus the integer $n$ extracted from quantum oscillations. The absolute value of the intercept of the linear fit yields the Berry phase, expected to be zero for conventional metals and 1/2 for Dirac Fermions. The Berry phase and the $\sim$ 1/2 intercept of the linear fit of LLs have been observed in Dirac materials, such as monolayer graphene and topological insulators.\cite{Nature 438 201} As shown in Fig. 3(d), the absolute intercept values of R$_{H}$ and R$_{ab}$ are 0.3 and 0.2, respectively, consistent with possible larger contribution of parabolic bands at the Fermi surface when compared to SrMnBi$_{2}$ and CaMnBi$_{2}$ or significant electron-hole asymmetry.\cite{PRB 84 220401,WangKF2,PRB 87 245104,Ando}

\begin{figure}
\centerline{\includegraphics[scale=0.3]{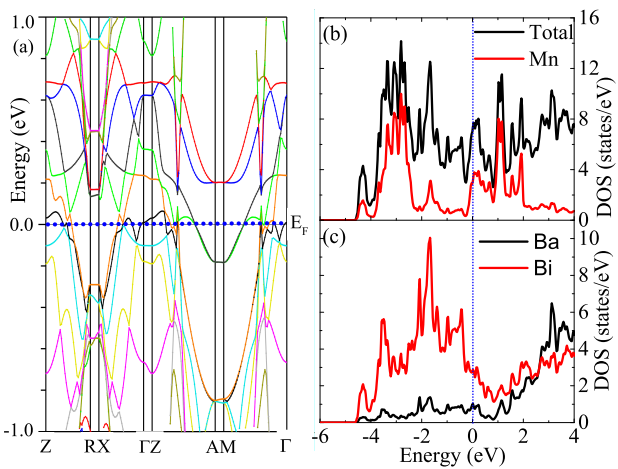}} \vspace*{-0.3cm}
\caption{Color online) . (a) The band structure for BaMnBi$_{2}$ with antiferromagnetic order but without spin-orbit coupling. (b,c) The total and local density of states from Mn, Ba and Bi atoms for AFM BaMnBi$_{2}$. The dotted line indicates the position of the Fermi energy. In the calculation, the moments of Mn atoms are arranged in an antiferromagnetic order in the ab plane but the Mn layers are ferromagnetically stacked along the c-axis.}
\label{XRD}
\end{figure}

The fast Fourier transform (FFT) spectrum of the resistivity oscillations at 1.55 K [Fig. 4(a)] reveals a periodic behavior in 1/B, a clear main frequency $F$ = 33.3 T and a possible weakly resolved frequency at 83.2T. Both are smaller than that of SrMnBi$_{2}$ ($F$ = 138 T) and CaMnBi$_{2}$ ($F_{1}$ = 101 T and $F_{2}$ = 181 T) (Table 1).\cite{PRB 84 220401,PRL 107 126402} The frequency of main oscillation is related to the Fermi surface averaged Fermi wave vector via the Onsager relation F=($\hbar$c/2$\pi$e)$\pi$k$_{F}$$^{2}$, giving  k$_{F}$ = 3.2$\times$10$^{8}$ $m^{-1}$ and the averaged carrier density n=[2/(2$\pi$)$^{3}$](4$\pi$/3k$_{F}$$^{3}$) = 1.10$\times$10$^{24}$m$^{-3}$.\cite{UK 1984,Eto,Ando} The cyclotron effective mass $m^{\ast}$ can be obtained by using the Lifshitz-Kosevitch formula:\cite{UK 1984} $\Delta$R$ = \frac{2\pi^{2}\kappa_{B}m^{\ast}T\gamma/eB\hbar}{\sinh(2\pi^{2}\kappa_{B}m^{\ast}T\gamma/eB\hbar}$, where, $\gamma = 1$ is a constant. Setting $K = \frac{2\pi^{2}\kappa_{B}m_{e}}{e\hbar}$ $\simeq$ 14.7TK$^{-1}$, $m^{\ast}=\mu\times$m$_{e}$,  B determined by the FFT range: 1/B = (0.03+0.09)/2 = 0.06. So $K/B \simeq 0.8824$. the above formula can be changed into the following simpler form:$\Delta$R/T$ = \frac{0.8824\mu}{\sinh(0.8824T\mu)}$. The fitting result is plotted by using the red line in Fig. 4(b) giving the cyclotron mass $m^{\ast} \approx 0.105m_{e}$, where m$_{e}$ is the free electron mass. The cyclotron mass $m^{\ast}$ of BaMnBi$_{2}$ is smaller than that of CaMnBi$_{2}$ with $m^{\ast} = 0.35m_{e}$ and SrMnBi$_{2}$ with $m^{\ast} = 0.29(2)m_{e}$ as shown in Table 1.\cite{WangKF2,PRL 107 126402}

Assuming that cyclotron mass reflects mostly quasi-2D Dirac fermions we estimate Fermi velocity from v$_{F}$=$\hbar$k$_{F}$/m$^{\ast}$=3.5$\times$10$^{5}$m/s where m$^{\ast}$=0.1m$_{e}$. The low magnetic field range of the in-plane magnetoresistance ($MR_{ab}$) unperturbed by quantum oscillations is shown in Figure 4(c). The experimental data can be well fitted by using semi-classical model, as $MR$ $\sim$ $A_{2}$$\cdot$$B^{2}$ for B$<$3T. In a multiband system with both Dirac and conventional parabolic-band carriers (such as electrons and holes), the coefficient of the low-field semi-classical $B^{2}$ quadratic term $A_{2}$ is related to the effective $MR$ mobility when the Dirac carriers are dominant in transport:$\sqrt{A_{2}}=\frac{\sqrt{\sigma_{e}\sigma_{h}}}{\sigma_{e}+\sigma_{h}}(\mu_{e}+\mu_{h})$ = $\mu_{MR}$, where $\sigma_{e}$, $\sigma_{h}$, $\mu_{e}$, $\mu_{h}$ are the effective electron and hole conductivity and mobility at zero field, respectively. Figure 4(d) shows the temperature dependence of $\mu_{MR}$. At about 2 K, the value of $\mu_{MR}$ is $\sim$ 3360 cm$^{2}$/Vs, similar to SrMnBi$_{2}$ with $\mu_{MR}$ is $\sim$ 3400 cm$^{2}$/Vs.\cite{PRB 84 220401}

Above a characteristic field $B^{\ast}$, as $B>B^{\ast}$, $MR$ is nearly linear $MR$$\sim$$A_{1}$$B$+$O$($B^{2}$), where $O$($B^{2}$) is very small. The evolution of $B^{\ast}$ with temperature is shown in Fig. 4(d). The energy splitting between the lowest and the first LLs of Dirac fermions can described by $\Delta_{LL}=\pm\nu_{F}\sqrt{2eB\hbar}$,
($\nu_{F}$ is the Fermi velocity).\cite{Nature 438 201, Science 324 924} In the quantum limit at specific temperature and field, $\Delta_{LL}$ becomes larger than both the Fermi energy $E_{F}$ and the thermal fluctuations $\kappa_{B}$$T$, so all carriers occupy the lowest Landau level, resulting in linear magnetoresistance.\cite{PRB 58 2788,Fundamentals} The critical field $B^{\ast}$ above which the quantum limit is satisfied at a specific temperature $T$ is $B^{\ast}=\frac{1}{2e\hbar\nu_{F}^{2}}(E_{F}+\kappa_{B}T)^{2}$.\cite{PRL 106 217004} The temperature dependence of the critical field $B^{\ast}$ in BaMnBi$_{2}$ can be well fitted by the above equation. The fitting [Fig. 5(b)] gives $E_{F}\approx7.38$meV, $\nu_{F}\approx1.61\times10^{5}$m/s (Table 1), in agreement with quantum oscillation measurement. The estimated Fermi velocity is similar to SrMnBi$_{2}$,\cite{PRB 84 220401} confirming the existence of Dirac fermion states in BaMnBi$_{2}$.

\bigskip
\begin{table*}[th]
\caption{Parameters for CaMnBi$_{2}$, SrMnBi$_{2}$ and BaMnBi$_{2}$. $T_{M}$ is the magnetic transition temperature, $m^{\ast}$ is the cyclotron mass, F is the frequency obtained by FFT, $E_{F}$ is the Fermion energy level, $\nu_{F}$ is the Fermi velocity, $\mu_{MR}$ is the effective $MR$ mobility.}
\begin{tabular}{cccccccc}
\hline\hline
Compounds & $T_{M}(K)$ & $m^{\ast}$($m_{e}$) & $F(T)$ & $E_{F}(meV)$ & $\nu_{F}(\times10^{5}m/s)$ & $\mu_{MR}(2K)(cm^{2}/Vs)$ & references \\ \hline
CaMnBi$_{2}$ & 270 & 0.35 & 101, 181 & $-$ & $-$ & 1800 & \cite{APL 100 112405, WangKF2}\\
SrMnBi$_{2}$ & 290 & 0.29(2) & 138 & 4.97 & 5.13 & 3400 &\cite{PRL 107 126402, PRB 84 064428, PRB 84 220401} \\
BaMnBi$_{2}$ & 290(1) & 0.105(1) & 33.3(4),83.2(5) & 7.37(3) & 1.60(8) & 3363(1) & this work  \\ \hline\hline
\end{tabular}
\end{table*}

The first principle calculation shows the presence of near-linear energy disperson and both electron- and hole-like bands near the Fermi level [Fig. 5(a)]. Both Mn and Bi-derived bands make contributions at the density of states (DOS) near E$_{F}$ [Fig. 5(b,c)]. Interestingly, we observe quantum oscillations in the Hall coefficient. Within Boltzmann transport theory, if electronic transport is governed by a single band with single scattering rate oscillations in $\rho$$_{xy}$ are not expected since the scattering rate cancels out in the Hall channel. This reveals the compensated nature of electronic transport with both electron and hole pockets at the Fermi surface in BaMnBi$_{2}$, i.e. at least two types of pockets with different mobility.\cite{PRB 87 245104,Kikugawa,Chakravarty} The positive sign indicates that the larger (or higher mobility) pocket corresponds to the hole band near $\Gamma$ point.\cite{PRB 87 245104}

Compensated nature of electronic transport is similar to not only another Dirac semimetal Cd$_{3}$As$_{2}$ but also to copper oxides and $\beta$-FeSe.\cite{HELP,Nicholas,WatsonMD} For BaMnBi$_{2}$ it indicates the presence of trivial (parabolic-like) electron band in addition to Dirac-like hole band at the Fermi surface,\cite{Ando} in agreement with Fig 5(a). On the other hand in SrMnBi$_{2}$ the electron-hole asymmetry due to spin-orbit coupling is intrinsic property of the anisotropic Dirac cone and therefore Berry phase cannot be quantized to $\beta$ = $\pi$.\cite{PRB 87 245104} This could also influence the intercept value shown in Fig. 4(d) and could contribute to quantum oscillations in the Hall channel.

\section{Conclusion}

In summary, we demonstrate the quantum transport in BaMnBi$_{2}$. When compared to other AMnBi$_{2}$ (A=Ca,Sr,Ba) materials, it can be concluded that alkaline earth atoms have minor influence on the antiferromagnetic transition temperature. Larger A atoms do bring the change of the unit cell space group from P4/nmm (A=Ca) to I4/mmm (A=Sr,Ba). The change of the space group, however, does change stacking of alkaline earth atoms above and below Bi square net from coincident (A=Ca) to staggered (A=Sr,Ba).\cite{PRB 87 245104} Somewhat elongated c-axis for A=Ba when compared to A=Sr material results in larger electron-hole asymmetry and deviation from the Berry phase $\beta$ = $\pi$. This agrees with first-principles result that the main factor for the anisotropic Dirac band is the anisotropic potential created by the Sr/Ba atoms.\cite{PRB 87 245104} AMnBi$_{2}$ materials therefore represent a platform to tune the anisotropy of Dirac cone states in the background of magnetic order. To that regard, it would be of interest to study magnetic exchange in BaMnBi$_{2}$ and its potential coupling to Dirac states.\cite{GuoYF}

\begin{acknowledgments}
Work at Brookhaven is supported by the U.S. DOE under Contract No. DE-AcO2-98CH10886. Work at the National High Magnetic Field Laboratory is supported by the NSF Cooperative Agreement No. DMR-0654118, and by the state of Florida
\end{acknowledgments}

\dag Present address: Key Laboratory of Materials Physics, Institute of Solid State Physics, Chinese Academy of Sciences, Hefei 230031, Peoples Republic of China.

$^{\ast }$Present address: Department of Physics, University of Maryland, College Park, MD 20742-4111, USA.

\end{document}